\documentclass[twocolumn,floatfix,prb,aps,showpacs,superscriptaddress]{revtex4-1} %,nofootinbib
\bibpunct{}{}{,}{n}{}{}

\usepackage{graphicx,epsf,amsmath,amssymb}
\usepackage{dcolumn}% Align table columns on decimal point
\usepackage{bm,bbm}% bold math 
\usepackage{enumerate}   
\usepackage[usenames,dvipsnames,svgnames,table]{xcolor}
\usepackage{subfigure}   

\usepackage{hyperref}
\hypersetup{
    bookmarks=false,         % show bookmarks bar?
    unicode=false,          % non-Latin characters in Acrobat's bookmarks
    pdftoolbar=true,        % show Acrobat's toolbar?
    pdfmenubar=true,        % show Acrobat's menu?
    pdffitwindow=false,     % window fit to page when opened
    pdfstartview={FitH},    % fits the width of the page to the window
    pdftitle={Dielectric and electronic properties of three-dimensional Luttinger semimetals with a quadratic band touching},    % title
    pdfauthor={UMontreal},     % author
    pdfsubject={Article},   % subject of the document
    pdfcreator={UMontreal},   % creator of the document
    pdfproducer={UMontreal}, % producer of the document
    pdfkeywords= {self} {energy} {luttinger} {semimetal}, % list of keywords
    pdfnewwindow=true,      % links in new window
    colorlinks=true,       % false: boxed links; true: colored links
    linkcolor=blue,          % color of internal links
    citecolor=red,        % color of links to bibliography
    filecolor=magenta,      % color of file links
    urlcolor=cyan           % color of external links
}

\begin{document}

\preprint{APS/123-QED}

\title{Dielectric and electronic properties of three-dimensional Luttinger semimetals with a quadratic band touching}

\author{S. Tchoumakov}  
\affiliation{D\'epartement de Physique, Universit\'e de Montr\'eal, Montr\'eal, Qu\'ebec, H3C 3J7, Canada}

\author{W. Witczak-Krempa}
\affiliation{D\'epartement de Physique, Universit\'e de Montr\'eal, Montr\'eal, Qu\'ebec, H3C 3J7, Canada}
\affiliation{Centre de Recherches Math\'ematiques, Universit\'e de Montr\'eal; P.O. Box 6128, Centre-ville Station; Montr\'eal (Qu\'ebec), H3C 3J7, Canada}
\affiliation{Regroupement Qu\'eb\'ecois sur les Mat\'eriaux de Pointe (RQMP)}

\date{\today}% It is always \today, today,

\begin{abstract}
Due to strong spin-orbit coupling, charge carriers in three-dimensional quadratic band touching Luttinger semimetals have non-trivial wavefunctions characterized by a pseudospin of 3/2. We compute the dielectric permittivity of such semimetals at finite doping, within the random phase approximation. Because of interband coupling, the dielectric screening shows a reduced plasma frequency and an increased spectral weight at small wavevectors, compared to a regular quadratic band. This weakens the effective Coulomb potential, modifying the single-particle self-energy for which we present both analytical and numerical results. At a low carrier density, the quasiparticle properties of this model strongly deviate from that of a single quadratic band. We compare our findings with experimental results for $\alpha-$Sn, HgSe, HgTe, YPtBi and Pr$_2$Ir$_2$O$_7$.  
\end{abstract}

\maketitle

\section{Introduction}

The interacting electron gas plays a fundamental role in our understanding of metals and semiconductors [\cite{rice,hedin,lundqv1,lundqv2,lundqv3}].  
It has been studied mostly in the case of a simple spin-degenerate quadratic band, where one can safely ignore the effects of the adjacents bands. 
However, this description is not sufficient to characterize Dirac and Weyl semimetals [\cite{Vafek14,reviewwilliam,reviewwsmlut}], where linearly dispersing conduction and valence bands meet at a point. The corresponding charge carriers have a non-trivial Berry phase that leads to many novel phenomena. Such physics is not restricted to linearly dispersing fermions and can also occur at quadratic band touching points [\cite{reviewwsmlut,topoqbt,savary,tpsmexp,blg1,blg2}]. In three dimensions, the latter can in fact be seen as a parent phase of a Weyl semimetal [\cite{WK12,WK13,balentsbaek,Savary14,iridatespiniceth}]. Materials with strong spin orbit coupling constitute a fruitful platform for the realization of such non-trivial semimetals or metals due to their tendency for band inversion [\cite{tpsmexp}]. When combined with strong electron correlations, found in materials with 4$d$ or 5$d$ active orbitals such as the iridium oxides, one is left with a rich playground for new physics that encompasses unconventional magnetism, superconductivity, quantum criticality, and even fractionalization [\cite{reviewwilliam,Vafek14,Maciejko15}].  

In this work, we study the dielectric and electronic properties of three-dimensional (3D) quadratic band touching Luttinger semimetals at finite doping [\cite{luttinger}]. This is motivated by experimental results for $\alpha-$Sn [\cite{graytinstruct,graytinarpes}], HgSe [\cite{hgsem1,hgsem2}], HgTe [\cite{hgtese,hgtebosc}], YPtBi [\cite{hfheuslerabinit,hfheuslerabinit2}] and Pr$_2$Ir$_2$O$_7$ [\cite{pr2ir2o7th,pr2ir2o7,iridatespiniceexp1,iridatespiniceexp2}]. Some optical [\cite{w01,w02,w03,w04,w05,q02}] and screening [\cite{q01,q02}] properties of Luttinger semimetals have been studied, \emph{i.e.}\ involving the dielectric permittivity $\epsilon(\omega,q)$ at either $q = 0$ or $\omega = 0$. Also, a screened Coulomb interaction in the absence of doping was shown to lead to non-Fermi liquid behavior [\cite{abrikosov,abrikosov2,balentsbaek}] which can transition to an interaction-driven topological insulator [\cite{mottluttinger1,mottluttinger2}]. However, this non-Fermi liquid regime has proven difficult to observe experimentally, even in Luttinger semimetals with a low carrier density such as the pyrochlore iridate Pr$_2$Ir$_2$O$_7$ [\cite{pr2ir2o7}]. We work in the complementary regime of small temperature at finite doping, where Fermi liquid behavior prevails, albeit with modifications coming from the non-trivial nature of the quadratic band touching. 

In Sec.~\ref{sec:screening}, we obtain the dielectric permittivity in the random phase approximation (RPA), and in Sec.~\ref{sec:self} we compute the consequences on the electronic self-energy. The interband transitions increase the screening of the long-range Coulomb potential and thus lead to a smaller renormalization of quasiparticle properties compared to a single quadratic band. We illustrate this with a numerical calculation of the quasiparticle residue, the effective mass and the compressibility of the electron gas. Finally, we compute the single-particle spectral function which shows plasmaron branches [\cite{lundqv3,diracplasmaron}] due to electron-plasmon coupling, and which are similar to replica bands due to electron-phonon coupling. In Sec.~\ref{sec:discussion}, we end with a comparison of our findings with experimental results on candidate materials for Luttinger semimetals.  We conclude in Sec.~\ref{sec:conclusion} by summarizing our main findings, and outlining some avenues for future research.

\begin{figure*}
    \centering
    \includegraphics[width = \textwidth]{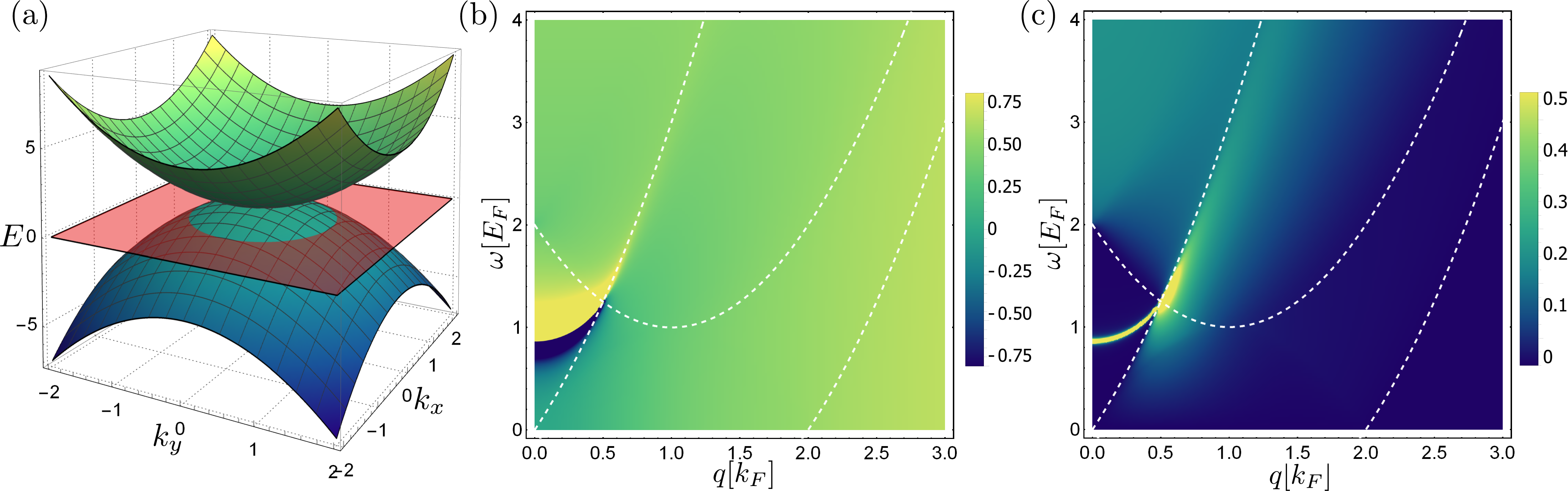}
    \caption{(a) Band-structure for Luttinger's model, the red plane is at the chemical potential. The upper and lower bands can be labelled by the helicity eigenvalues $\lambda = \pm 3/2$ and $\pm 1/2$, respectively. (b-c) The (b) real and (c) imaginary parts of the inverse dielectric permittivity, $1/\epsilon(\omega,q)$, for $r_s = 2$ as a function of wavevectors, $q$, and (real) frequencies, $\omega$. The white dashed lines indicate the branches of the particle-hole excitation diagram.}
    \label{fig:spctrmeps}
\end{figure*}

\section{Model} 
In Luttinger's model [\cite{luttinger}], the effective non-interacting Hamiltonian for the four bands is
\begin{align}\label{eq:h0}
    \hat{H}_0 = \frac{\hbar^2}{2m} \left[ (\alpha_1  - 5\alpha_2/4) {\bf k}^2  + \alpha_2 \left({\bf k}\cdot \hat{{\bf J}}\right)^2 \right] - \mu.
\end{align}
Here, we denote the band mass $m$ and the $j = 3/2$ total angular momentum operators $\hat{{\bf J}} = (\hat{J}_x,\hat{J}_y,\hat{J}_z)$. This model has full rotational symmetry and helical symmetry, defined by the helicity operator $\hat{\lambda} = {\bf k}\cdot \hat{{\bf J}}/k$ [\cite{schliemann,savary}]. The eigenstates of Eq.~\eqref{eq:h0} can be labeled by the eigenvalues $\lambda = \pm 1/2,\pm 3/2$ of the helicity operator and the corresponding spectrum is $\varepsilon_{\pm}({\bf k}) = \pm\hbar^2 k^2/2m_{\pm}$ with $m_{\pm} = m/(\alpha_2 \pm \alpha_1)$. In the following we 
assume that the bands are particle-hole symmetric, \emph{i.e.}\ $\alpha_1 = 0$ and $\alpha_2 = 1$, and take $\mu < 0$ for a hole Fermi surface (see Fig.~\ref{fig:spctrmeps} (a)). We analyze the effect of particle-hole asymetry in Sec.\ \ref{sec:discussion} and in Appendix~\ref{app:plasma}.

\section{Screening properties}
\label{sec:screening}

The dielectric permittivity of Luttinger's model with Coulomb repulsion between electrons in a neutralizing background is, within the RPA,
\begin{align}
    \label{eq:dielpol}
    \epsilon_{\rm RPA}(i\Omega,{\bf q}) = 1 - V({\bf q})\Pi(i\Omega,{\bf q}),
\end{align}
where the bare electric potential is, in CGS units, 
\begin{align}
    V({\bf q}) = 4\pi e^2/(\epsilon^* q^2)
\end{align}
with $\epsilon^*$ the background dielectric permittivity. Also, the charge polarizability is
\begin{align}
    \label{eq:polarizability}
    &\Pi(i\Omega,{\bf q}) = \\
    &\sum_{\sigma \sigma^{\prime} \bf p} \frac{f_{D}(\varepsilon_{\sigma}({\bf p})) - f_{D}(\varepsilon_{\sigma^{\prime}}({\bf p}+{\bf q}))}{i\Omega +\varepsilon_{\sigma}({\bf p}) - \varepsilon_{\sigma^{\prime}}({\bf p}+{\bf q})} {\rm Tr}\left[\hat{P}_{\sigma}({\bf p}) \hat{P}_{\sigma^{\prime}}({\bf p}+{\bf q})\right].\nonumber
\end{align} 
These expressions are written in terms of the imaginary frequency $i\Omega$ and one obtains the analytic continuation to the real-frequency axis for $i\Omega \rightarrow \omega + i0^+$ [\cite{quinnferrel}]. The indices $\sigma = \pm$ refer to upper and lower bands with energy $\varepsilon_{\sigma}({\bf k}) = \sigma \hbar^2k^2/2m$ and projector $\hat{P}_{\sigma}({\bf k}) = \frac12\left[ \hat{\mathbbm{1}} + \hat{H}_0({\bf k})/\varepsilon_{\sigma}({\bf k})\right]$. Since we perform calculations at zero temperature, the Fermi distribution reduces to $f_D(\varepsilon) = \Theta(E_F - \varepsilon)$, with $E_F$ the Fermi energy.  In all what follows we set $\hbar = 1$, such that time and energy scales are in units of the Fermi energy, $E_F$, and  length and wavevectors are in units of the Fermi wavevector, $k_F$. This choice of units allows us to write all expressions as a function of the Wigner-Seitz radius, $r_s = m e^2/\alpha\epsilon^* k_F$ with the constant $\alpha = (4/9\pi)^{1/3} \approx 0.52$, $k_F = (3\pi^2 n)^{1/3}$ and where $E_F = -k_F^2/2m < 0$ is in the bottom band. The band structure is particle-hole symmetric and our observations are thus independent on the sign of the Fermi energy.

The charge polarizability \eqref{eq:polarizability} explicitly depends on the eigenspinor overlaps through 
\begin{align}
    {\rm Tr}\left[\hat{P}_{\sigma_1}({\bf k}+{\bf q}) \hat{P}_{\sigma_2}({\bf k})\right] = \frac12\left\{ 2 + \sigma_1\sigma_2 \left[ 3 \cos^2(\theta_{{\bf k} + {\bf q}, {\bf k}}) - 1\right] \right\},\nonumber
\end{align}
with $\theta_{{\bf k} + {\bf q}, {\bf k}}$ the angle between ${\bf k} + {\bf q}$ and ${\bf k}$. This overlap function is a key ingredient in the description of narrow bandgap semiconductors [\cite{smallgap1}]. It does not only allow for interband transitions but also affects scattering within the same band, as discussed in the context of superconductivity with spin-orbit interaction~[\cite{socsc2,tpsmexp,savary,venderbos,superluttinger}].

The analytic expression of polarizability is established for a single quadratic band [\cite{giuliani}], for Dirac bands [\cite{2ddirac,3ddirac}] and for hole bands in zinc-blende semiconductors [\cite{zcbldhole}]. Its expression for a quadratic band touching described by Luttinger model in Eq.~\eqref{eq:h0} was considered either in the static or long-range limits [\cite{w01,w02,w03,w04,w05,q02,q01,q02}], and we derive its expression for all wavevectors and frequencies in Appendix~\ref{app:pintrainter}. There we decompose polarizability in Eq.~\eqref{eq:polarizability} into intraband and interband contributions, $\Pi = \Pi^{\rm intra} + \Pi^{\rm inter}$. These two expressions depend on the following integrals
\begin{align}
    &I_{+}(a,b; \alpha,\beta) = \int_{a}^{b} dx~ \frac{\log(x+\alpha)}{x+\beta}\\
    &I_{-}(a,b; \alpha,\beta) = \int_{a}^{b} dx~ \frac{\log(|x-\alpha|)}{x+\beta},
\end{align}
which also appear in the context of hole screening in zinc-blende semiconductors [\cite{zcbldhole}] and in quark physics [\cite{loopintegral}]. The expressions of $I_{\pm}$ are given in Eq.~\eqref{eq:Ipmexp} of Appendix~\ref{app:pintrainter} and the resulting components of polarizability are 
\begin{align}
    \begin{split}
    \Pi^{\rm intra}(i\Omega,q) = &\frac2q \Psi_{\rm Lindhard}(i\Omega,q) + \Psi_2(i\Omega,q)\\
    & + (i\Omega \rightarrow -i\Omega),
    \end{split}\\
    \Pi^{\rm inter}(i\Omega,q) = &\Psi_3(i\Omega,q) + (i\Omega \rightarrow -i\Omega),    
\end{align}
which we write in terms of the Lindhard function $\Psi_{\rm Lindhard}(z = \frac{i\Omega}{2q} - q/2) = N_0\left(\frac{z}{2} + \frac{1-z^2}{4}\ln\left(\frac{z+1}{z-1}\right)\right)$ [\cite{giuliani}], with $N_0 = 1/(4\pi^2)$ the density of states per band at the Fermi surface, and the following two functions $\Psi_2(q,i\Omega)$ and $\Psi_3(q,i\Omega)$ (see Appendix~\ref{app:pintrainter})
\begin{widetext}
\begin{align}
    \Psi_2(i\Omega,q) &= -\frac{3qN_0}{8}\left\{ \left( 1 + \frac{z^2}{i\Omega} \right) \sum_{\substack{\sigma,\tau = \pm}} \sigma I_{+}(0, \sigma ; z, \tau i \sqrt{i\Omega}) + \frac12\left( 2 + \frac{1}{u} + u \right) \sum_{\substack{\sigma,\tau = \pm}} \sigma I_{\sigma}(0, 1/q ; 1, \tau i\sqrt{u}) \right. \nonumber\\
    &\left. - \frac{1}{u}\sum_{\sigma = \pm} \sigma I_{\sigma}(0, 1/q ; 1, 0^+) + \frac12 \left( 1 - \frac{1}{q^2} \right) \log\left( \frac{1+q}{|1-q|} \right) - 2\frac{z^2}{i\Omega}\left( {\rm Li}_2(1/z) - {\rm Li}_2(-1/z)\right) + \frac1q\right\},
    \label{eq:psi2}
\end{align}
\begin{align}
    \Psi_3(i\Omega,q) &= \frac{3N_0}{16q}\left\{ 2\sum_{\beta = \pm} (z_{\beta}^2 - 1) \log\left( \frac{1- z_{\beta} }{1+z_{\beta}} \right) - 2 q - (1-q^2) \log\left( \frac{1+q}{|1-q|} \right) + \frac{2 q^4}{i\Omega} \left( \sum_{\sigma = \pm} \sigma I_{\sigma}(0,1/q ; 1, 0^+) - \pi^2/2\right) \right.\nonumber \\
    &\left.+ \frac{(q^2 - i\Omega)^2}{i\Omega} \sum_{\sigma = \pm} \sigma \left[ 2 I_{\sigma}(0,q ; 1, 0^+) - \sum_{\tau = \pm}\left( I_{\sigma}(0,q;1,\tau/\sqrt{u})  - \frac12 \sum_{\beta = \pm} I_{+}(0,-\sigma ; -1/z_{\beta}, \tau/\sqrt{i\Omega}) \right) \right]\right\},
    \label{eq:psi3}
\end{align}
\end{widetext}
with $z = i\Omega/(2q) - q/2$, $u = i\Omega/q^2$ and $z_{\pm} = q/2 \pm \sqrt{i\Omega/2 - q^2/4}$.

The contribution $\Psi_2$ describes the influence of the eigenspinor overlap on intraband polarizability while $\Psi_3$ describes interband transitions. The behaviour of the corresponding dielectric function is computed on the real axis with the analytic continuation $i\Omega \rightarrow \omega + i0^{+}$ and we show the behaviour of $1/\epsilon(\omega,q)$ over real frequencies in Figs.~\ref{fig:spctrmeps}(b,c).

\begin{figure} 
    \centering
    \includegraphics[width = 0.45\textwidth]{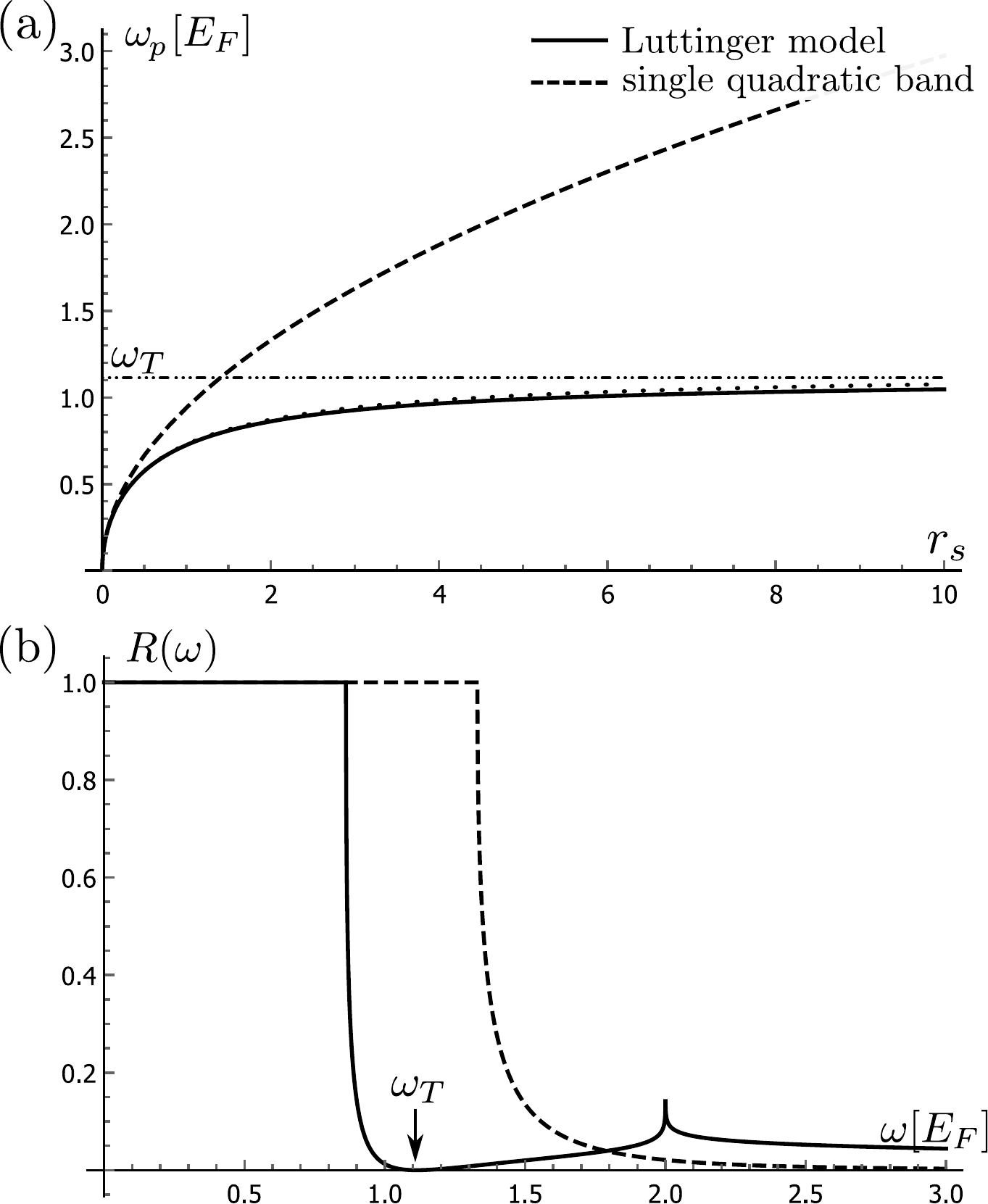}
    \caption{(a) Plasma frequency for the single quadratic band (dashed line) and for Luttinger's model (plain line) as a function of the Wigner-Seitz radius. The dotted line is the approximate expression of the plasma frequency Eq.~\eqref{eq:wpapprox} and the dot-dashed is the transparency frequency, $\omega_T = 1.113 \mu$. (b) Normal incidence reflectivity as a function of frequency for the single quadratic band (dashed line) and for Luttinger's model (plain line) for $r_s = 2$. The electron gas does not transmit light below the plasma frequency ($R=1$) and is transparent for $\omega = \omega_T$. The reflectivity peaks at the onset of interband transition, for $\omega = 2 E_F$.}
    \label{fig:wprs}
\end{figure}

The real part of the dielectric permittivity (see Fig.~\ref{fig:spctrmeps}(b)) is close to that for a single quadratic band and changes sign at the plasma frequency (see below). At large wavevector, the screened Coulomb potential contains a term linear in wavevector $q$. For comparison with the literature [\cite{abrikosov}], we write its expression in dimensioned units and for $q^2/2m \gg \omega, E_F$
\begin{align}
    \frac{V(q)}{\epsilon(\omega,q)} \approx \frac{4\pi e^2}{\epsilon^* q^2 + m e^2 q a},
\end{align}
with $a \approx 1.712$. This linear increase of the polarizability would dominate for a large Wigner-Seitz radius, for $k_F \ll q \ll me^2/\epsilon^*$, and was first discussed by Abrikosov and Beneslavski [\cite{abrikosov}] in the context of quadratic band touching without doping. There, the Fermi surface is reduced to a point with a non-Fermi liquid behaviour in presence of the Coulomb potential [\cite{abrikosov,abrikosov2,balentsbaek}]. 
%A similar increase of the charge polarizability for large wavevectors is also obtained for 2D Dirac but can be absorbed as an effective background dielectric permittivity because of dimensionality [\cite{2ddirac}].

The imaginary part of the dielectric permittivity Fig.~\ref{fig:spctrmeps}(c) shows the expected particle-hole excitation diagram, denoted by white dashed lines, with (1) intraband excitations for $\omega_{+}(q) = q^2 + 2q \geq \omega \geq \omega_{-}(q) = q^2 - 2q$ and (2) interband excitations for $\omega > \omega_{12}(q) = 1 + (1 - q)^2$. The plasmon branch is gapped and unlike Weyl semimetals [\cite{chargespin}], it does not carry spin because of the compensating helicity on each band (see Appendix~\ref{app:wpnospin}). Also, the plasma frequency vanishes in absence of doping but may reappear due to thermal fluctuations at non-zero temperature~[\cite{plasmonsthermal}]. The approximate plasma frequency is (see Appendix~\ref{app:plasma})
\begin{align}
    \label{eq:wpapprox}
    \omega_p \approx \sqrt{\frac{16 \alpha r_s}{3\pi(1+4\alpha r_s/\pi)}}.
\end{align}
In the limit of a small Wigner-Seitz radius the plasma frequency is that of a single quadratic band [\cite{giuliani}] while at large $r_s$ it saturates to the frequency $\omega_T \approx 1.113 E_F$, as illustrated in Fig.~\ref{fig:wprs}(a). A similar decrease in the plasma frequency was discussed in the context of Dirac semimetals [\cite{quantplasma}]. This is a consquence of the increase in the dielectric function by interband excitations, such that the reflectivity of the electron gas (see Fig.~\ref{fig:wprs}(b))
\begin{align}
    R(\omega) = \left| \frac{\sqrt{\epsilon(\omega,q = 0)} - 1 }{\sqrt{\epsilon(\omega,q = 0)} + 1 } \right|^2
\end{align}
vanishes at $\omega = \omega_T$ with $\lim_{q\rightarrow 0} \Pi(\omega_T,q)/q^2 = 0$. The transparency frequency $\omega_T$ is independent of the Wigner-Seitz radius because it only depends on the charge polarizability and not on the Coulomb potential. This transparency window lies between the plasma frequency and the onset of interband transitions at $2E_F$ and its experimental observation depends on the background dielectric permittivity, $\epsilon^*$.

Finally, in the dynamic regime, $\omega \gg q^2$, we retrieve the optical dielectric function discussed in Refs.\ [\cite{q01,w05,pr2ir2o7}] as explained in Appendix~\ref{app:plasma}. Its behaviour for large frequencies, $\omega \gg 2E_F$, is $\epsilon(\omega) \sim \epsilon^* + 2 e^{i\pi/4} \sqrt{E_0/\omega}$ where the characteristic energy $E_0/E_F = \alpha^2 r_s^2$ is the excitonic energy. We thus have an increased screening of the long-range Coulomb potential for larger frequencies, and in Sec.\ \ref{sec:renorm} we show how it affects the self-energy as well as quasiparticles properties of the Luttinger semimetal.

\section{Self-energy}
\label{sec:self}

The one-particle self-energy is diagonal in the eigenvector basis, $\hat{\Sigma}(i\Omega,{\bf q}) = \sum_{\sigma}\Sigma_{\sigma}(i\Omega,{\bf q})\hat{P}_{\sigma}({\bf q})$. We decompose it into an exchange and a correlation part $\Sigma_{\sigma}(i\Omega,{\bf q}) = \Sigma^{\rm (ex)}_{\sigma}({\bf q}) + \Sigma^{\rm (c)}_{\sigma}(i\Omega,{\bf q})$ such that the exchange self-energy corresponds to the Hartree-Fock self-energy, which is independent of frequency,
\begin{align}
    \begin{split}
    \Sigma^{\rm (ex)}_{\sigma}({\bf q}) = \frac{-1}{2\mathcal{V}}\sum_{\sigma^{\prime} {\bf k} } f_{D}(\varepsilon_{\sigma^{\prime}}&({\bf q}-{\bf k})) V({\bf k})\\
    &\times {\rm Tr}\left[\hat{P}_{\sigma}({\bf q}) \hat{P}_{\sigma^{\prime}}({\bf q -k})\right],
    \end{split}
    \nonumber
\end{align}
with $\mathcal{V}$ the volume of the electron gas. We further decompose this contribution into an intrinsic and an extrinsic part. They respectively describe the situation of a vanishing chemical potential and the corrections due to a non-zero Fermi energy, $\Sigma_{\sigma}^{\rm (ex)} = \Sigma_{\sigma}^{\rm int} + \Sigma_{\sigma}^{\rm ext}$, which we compute by introducing a cutoff $\Lambda \gg k_F$ (see Appendix~\ref{sec:exse})
\begin{align}
    \label{eq:resint}
    \Sigma_{\sigma}^{\rm int} &= \frac{\alpha r_s}{\pi} \left( -2 \Lambda/k_F + \frac{3\sigma \pi^2 q}{16}\right),\\
    \begin{split}
        \Sigma_{\sigma}^{\rm ext} &= \frac{\alpha r_s}{32\pi} \left[ 16(2+\sigma) + 6\sigma\left( 1/q^2 - 1 \right)\right.\\
        & + \frac{(q^2-1)(3\sigma + q^2(7\sigma - 16))}{q^3} \log\left( \frac{1+q}{|1-q|} \right) \\
        & - 2\sigma q\left( \pi^2  + 6\left\{ \log\left( |1-q| \right)\log( q )- \log^2(q) \right.\right.\\
        &~~~~~~~~~~~~~~~~~~ \left.\left.\left.- {\rm Li}_{2}(-1/q) - {\rm Li}_{2}(1-1/q) \right\} \right) \right].
    \end{split}
\end{align}
The only dependence on the cutoff, $\Lambda$, is as a constant shift in the intrinsic part $\Sigma_{\sigma}^{\rm int}$ and can be absorbed in the chemical potential. This expression of the exchange self-energy would lead to a singular effective mass at $k = k_F$ for the filled band (here, $\sigma = -$). This singularity is due to the long-range nature of the bare Coulomb potential and also happens for a single quadratic band. It vanishes if one introduces the effect of screening in the self-energy, as described by the correlation self-energy
\begin{align}
    \begin{split}
    \Sigma_{\sigma}^{\rm (c)}(&i\nu,{\bf q}) = - \frac{1}{2\mathcal{V}}\int \frac{d\Omega}{2\pi} \sum_{\sigma^{\prime} {\bf k} } G_{\sigma^{\prime}}(i(\nu - \Omega),{\bf q}-{\bf k})\\
    &\times V({\bf k})\left( \frac{1}{\epsilon(i\Omega,{\bf k})} - 1 \right) {\rm Tr}\left[\hat{P}_{\sigma}({\bf q}) \hat{P}_{\sigma^{\prime}}({\bf q -k})\right],
    \end{split}
\end{align}
where the Green's functions of Luttinger's model are
\begin{align}
    G_{\sigma}(i\Omega,{\bf k}) = \frac{1}{-i\Omega + \xi_{\sigma}({\bf k})},
\end{align}
with $\xi_{\sigma}({\bf k}) = 1 + \sigma k^2$. The expression of the correlation self-energy on the real-frequency axis, using Feynman's prescription, can be decomposed as a sum of a line integral and a residue [\cite{quinnferrel,lundqv1,lundqv2,lundqv3,diracplasmaron}]
\begin{align}
    &\Sigma_{\sigma}^{\rm (c)}(\omega,{\bf q}) = \Sigma_{{\rm line},\sigma}(\omega, q) + \Sigma_{{\rm res},\sigma}(\omega, q),\\
    &\Sigma_{{\rm line},\sigma}(\omega,{\bf q}) = - \frac{1}{2\mathcal{V}}\sum_{\sigma^{\prime} {\bf k}} \int\frac{d\Omega}{2\pi}~ G_{\sigma^{\prime}}(\omega - i\Omega, {\bf q} - {\bf k})V({\bf k})\nonumber\\
    & ~~~~~ ~~~~~~\times\left[ \frac{1}{\epsilon(i\Omega,k)} - 1 \right]{\rm Tr}\left[\hat{P}_{\sigma}({\bf q}) \hat{P}_{\sigma^{\prime}}({\bf q -k})\right], \\
    \begin{split}
    &\Sigma_{{\rm res},\sigma}(\omega,{\bf q}) = \frac{1}{2\mathcal{V}}\sum_{\sigma^{\prime} {\bf k}} \left[ \Theta(\omega - \xi_{\sigma^{\prime}}({\bf q} - {\bf k})) - \Theta( - \xi_{\sigma^{\prime}}({\bf q} - {\bf k})) \right]V({\bf k})\\
    &\times\left[ \frac{1}{\epsilon\left[\left(\omega - \xi_{\sigma^{\prime}}({\bf q} - {\bf k})\right)\left( 1 - i0^+\right),{\bf k}\right]} - 1 \right]{\rm Tr}\left[\hat{P}_{\sigma}({\bf q}) \hat{P}_{\sigma^{\prime}}({\bf q -k})\right].
    \end{split}
\end{align}
It should be noted that the dielectric permittivity appears with $\epsilon(\omega,{\bf q}) = \epsilon(-\omega,{\bf q})$ since with our convention the two frequencies are on opposite sides of the branch cut [\cite{quinnferrel}].

In the following we present numerical results for the resulting self-energy. We first confirm that the the interacting Luttinger semimetal is a Fermi liquid and then investigate some of its quasiparticle properties, namely the quasiparticle weight, the renormalized mass and the compressibility. We finish with a discussion of the spectral function which, besides the renormalized electron branch, contains a plasmaron branch due to electron-plasmon coupling.

\begin{figure}
    \centering
    \includegraphics[width = 0.45\textwidth]{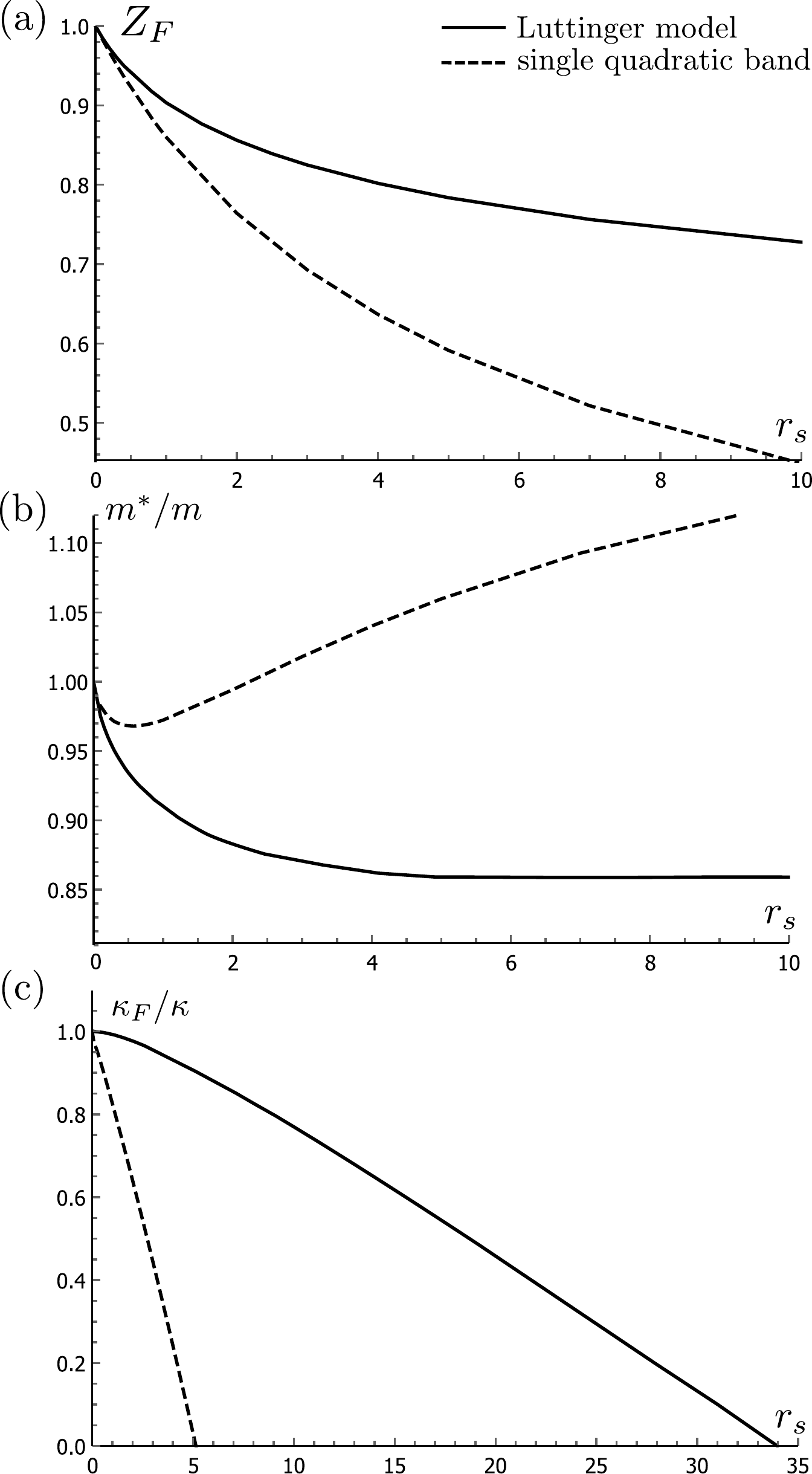}
    \caption{Quasiparticle properties for the single quadratic band (dashed line) and for Luttinger's model (plain line) as a function of the Wigner-Seitz radius. In (a) the quasiparticle weight $Z(\omega=0,k=k_F)$, in (b) the renormalized mass $m^*$ and in (c) the inverse compressibility $1/\kappa$. The difference between the two models is due to the absence or presence of interband coupling which stabilizes the electron gas for small carriers densities (large $r_s$).}
    \label{fig:oneparticle}
\end{figure}

\subsection{Quasiparticle lifetime}
\label{sec:lifetime}
The only contribution to the imaginary part of the self-energy comes from the residue part. The inverse quasiparticle lifetime is
\begin{align}
    \frac{1}{\tau(k)} &= -2 {\rm sign}[\xi_{-}(k)] {\rm Im}\left[ \Sigma_{-}(\xi_-(k),{\bf k}) \right] \\
   = - \frac{\alpha r_s}{\pi^2} &\int_{1}^{k} dq q^2\int_{-1}^{1} du~ \frac{1 + 3 u^2}{2}{\rm Im}\left[ \frac{ 1}{q^2 \epsilon(2(k-q), {\bf k -q})}\right].\nonumber
\end{align}
In the last expression we used the variable $u = \cos \theta$ and expanded $\xi_{-}(k) - \xi_{-}(q) \approx 2(q-k)$ for $k,q$ near the Fermi surface. In this limit we find the same expression as for a single quadratic band
\begin{align}
    \lim_{q \rightarrow 0}\lim_{\omega \rightarrow 0} {\rm Im}\left[ \frac{V(q)}{\epsilon(\omega,q)} \right] = \frac{\pi^2 \omega}{16 \alpha r_s q},
\end{align}
from which we deduce the quasiparticle lifetime of Luttinger's model near the Fermi surface
\begin{align}
    1/\tau(k) \approx 3(k-1)^2/20 \sim \xi_k^2.
\end{align}
The quasiparticle properties, such as the quasiparticle residue, the effective mass and compressibility, are thus well defined. We now compare these quantities to what is obtained for a single quadratic band [\cite{rice,hedin}].

\subsection{Quasiparticle properties}
\label{sec:renorm}

The residue $Z_F$ quantifies the spectral weight of quasiparticles :
\begin{align}
    Z_F = \frac{1}{1 - \lim_{(\omega,q)\rightarrow(0,1)} \frac{\partial}{\partial \omega} \Sigma_{-}(\omega,q)}.
\end{align}
This quantity does not decrease as much as for a single quadratic band, as illustrated in Fig.~\ref{fig:oneparticle}(a), as a consequence of the increased screening of the long-range part of the bare Coulomb potential by interband transitions. This larger quasiparticle residue alters other renormalized quantities, such as the effective mass, $m^*$
\begin{align}
    \frac{m}{m^*} = Z_F \left[ 1 - \lim_{(\omega,q)\rightarrow(0,1)}\frac{\partial \Sigma_{-}}{\partial q}(\omega,q)\right],
\end{align}
which decreases for all values of $r_s$, while it is non-monotonous for the single quadratic band (see Fig.~\ref{fig:oneparticle}(b)). We find that the effective mass decreases down to $m^*/m \approx 0.86$, that we have checked for $r_s$ up to $200$. The electron-electron repulsion thus decreases the quasiparticles density of states at large $r_s$ which in turn affects the electron gas screening properties, described by the compressibility $\kappa$.

The thermodynamic expression of compressibility is $1/\kappa = -\mathcal{V} (\partial P/\partial \mathcal{V})|_{N}$ where the pressure $P = -(\partial E/\partial \mathcal{V})|_{N}$ relates changes of the total energy $E$ with the volume $\mathcal{V}$. The total energy for a given carrier density, $n = k_F^3/(3\pi^2)$, is related to the chemical potential $\mu = E_F + {\rm Re}\left[ \Sigma_-(\omega = 0, k = k_F) \right]$ through the theorem of Seitz [\cite{mahan}] and allows us to use to following expression for compressibility $1/\kappa = n^2 d\mu/dn$.

We find that the compressibility of Luttinger semimetals is smaller than for a single quadratic band and we draw it in Fig.~\ref{fig:oneparticle}(c) in units of the free electron-gas compressibility $\kappa_F = 3/(2E_F n)$. The inverse compressibility vanishes for $r_s \approx 34$, a value much larger than that for a single quadratic band where this critical Wigner-Seitz radius is of $5.1$. Beyond this value the electron gas is unstable and this indicates that, because of interband coupling, zero gap semiconductors are more stable than metals for small carrier densities.

We note that the compressibility is independent on the cutoff wavevector, $\Lambda$, introduced in Eq.~\eqref{eq:resint}. Indeed, this term leads to a correction to the self-energy which is independent on the carrier density.

\subsection{Spectral function}

\begin{figure}[t!]
    \centering
    \includegraphics[width = 0.45\textwidth]{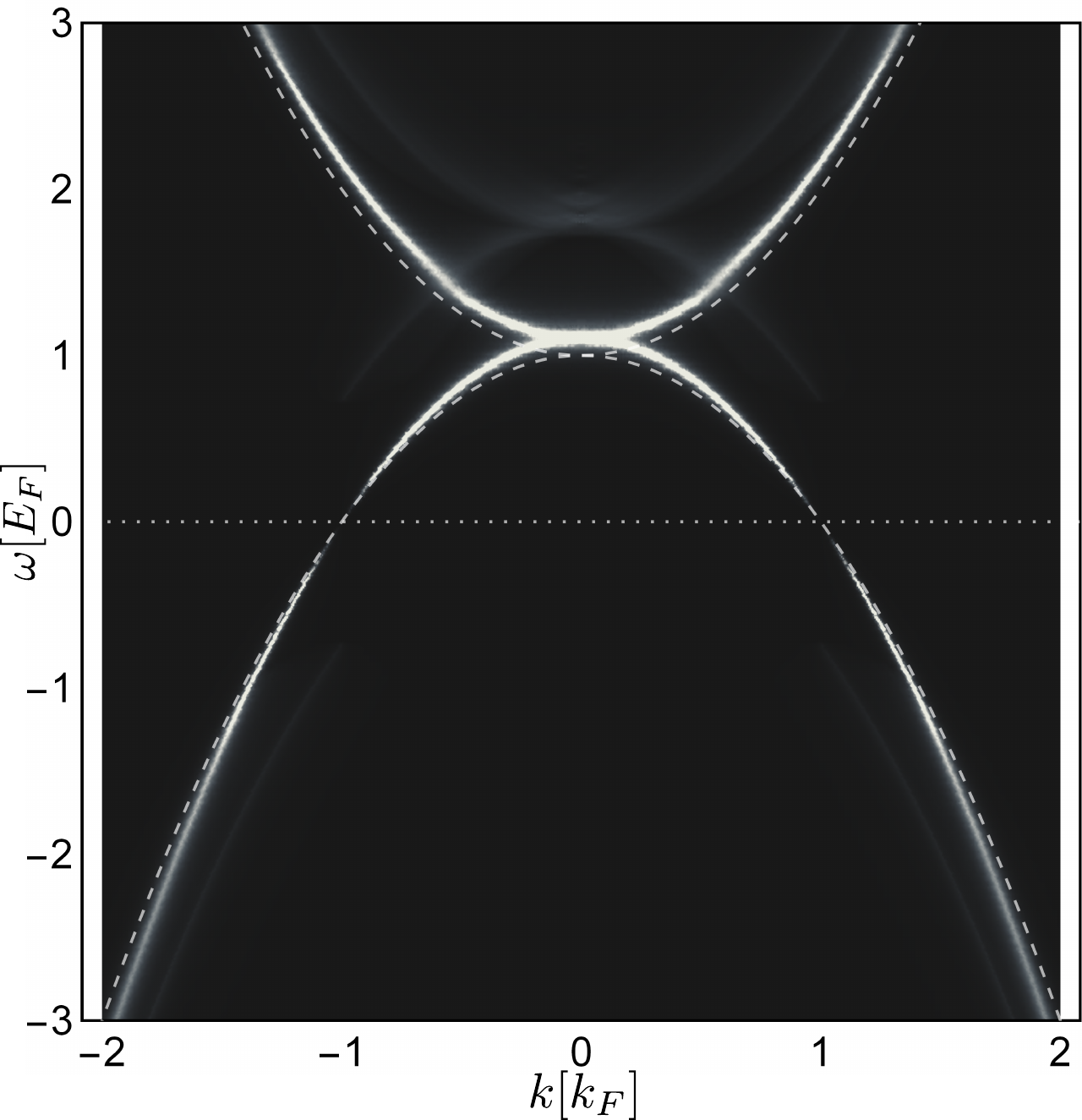}
    \caption{Spectral function of the interacting Luttinger's model for $r_s = 1$. The dashed line represents the non-interacting spectrum and it does not deviate much from the spectral function of the interacting Luttinger semimetal. We also observe plasmaron peaks due to the hybridization of electrons and plasmons. The dotted line locates the Fermi surface, at $\omega = 0$.}
    \label{fig:spctral}
\end{figure}

\begin{figure}[t!]
    \centering
    \includegraphics[width = 0.45\textwidth]{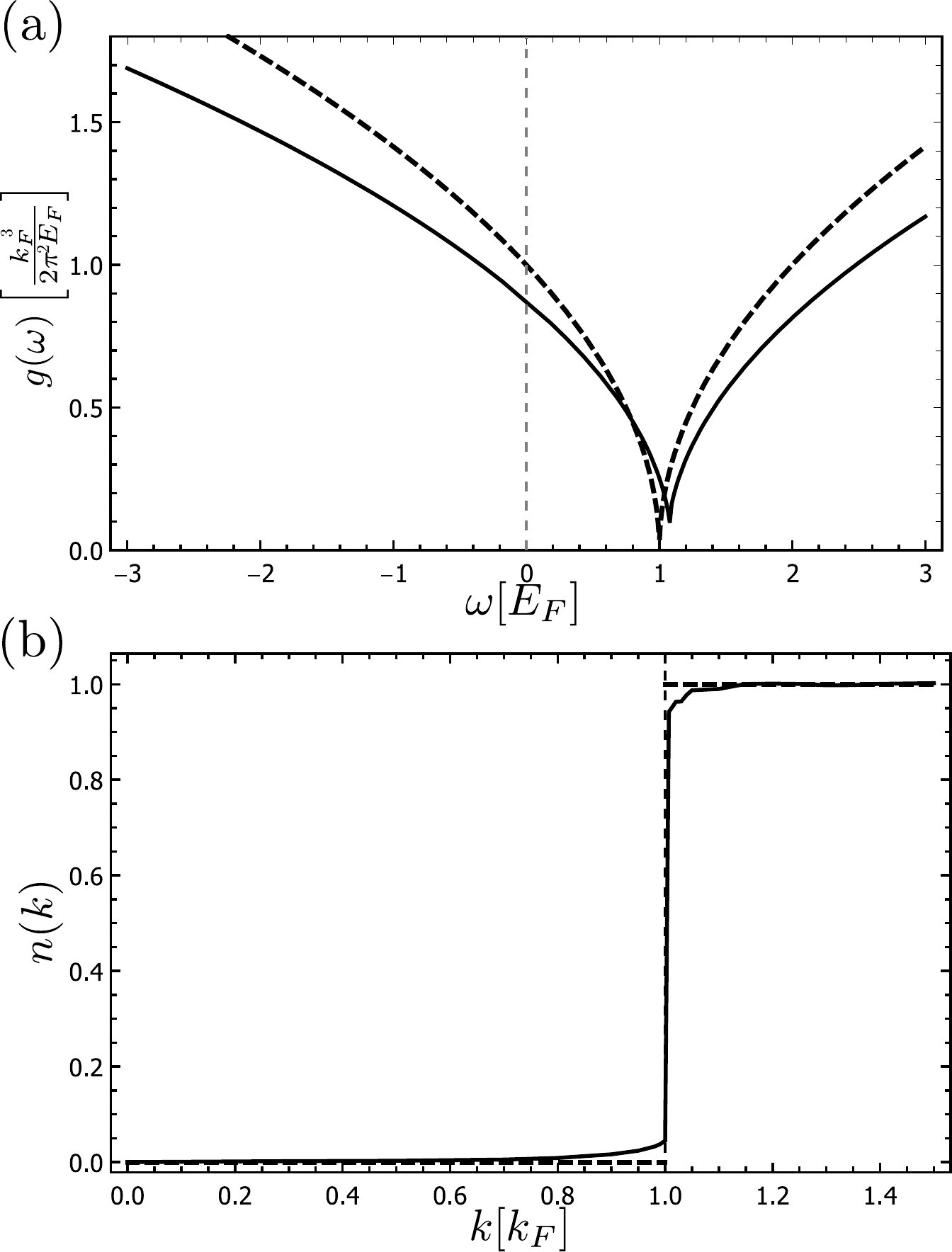}
    \caption{(a) Density of states and (b) filling of the non-interacting (dashed line) and interacting (plain line) Luttinger model for $r_s = 1$. The main contribution to the density of states comes from quasiparticles and is smaller at the Fermi surface because of the reduced effective mass. The filling function shows occupied states above the Fermi sea and some unoccupied states below it, the jump at $k = k_F$ equals the quasiparticle residue $Z_F$.}
    \label{fig:nk}
\end{figure}

In the Feynman prescription, the single-particle spectral function $A(\omega, q)$ is
\begin{align}
    A(\omega,q)& =\\
    &-\frac{1}{\pi}\sum_{\sigma} \frac{{\rm sign}(\omega){\rm Im}\Sigma_{\sigma}(\omega,q)}{(\omega - \xi_{\sigma}(q) - {\rm Re}\Sigma_{\sigma}(\omega,q))^2 + {\rm Im}\Sigma_{\sigma}(\omega,q)^2}\nonumber.
\end{align}
We illustrate its behaviour in Fig.~\ref{fig:spctral} for $r_s = 1$. The hole band acquires a finite width away from the Fermi surface due to electron-hole and plasmon excitations, in agreement with our expansion in Sec.\ \ref{sec:lifetime}. The satellite peaks correspond to plasmaron branches [\cite{lundqv3,diracplasmaron}] which result from the coherent electron-plasmon and hole-plasmon coupling. The bare dispersion is shown with dashed lines and we observe that the quasiparticle and plasmaron branches follow it closely, the shift of the plasmaron spectrum is essentially the plasma frequency in Fig.~\ref{fig:wprs}(a).

One can extract the density of states $g(\omega) = 2 \int d^3{\bf k}/(2\pi)^3 A(\omega,{\bf k})$ and the filling function $n(k) = \int_{-\infty}^{0}d\omega~ A(\omega,{\bf k})$ from the spectral function [\cite{lundqv3}]. We illustrate these functions in Fig.~\ref{fig:nk} for $r_s = 1$. We observe that the plasmaron contribution to the density of states is vanishingly small. Also, the density of states at the Fermi surface decreases in agreement with the decrease in the effective mass shown in Fig.~\ref{fig:oneparticle}(b). The filling function $n(k)$ has a discontinuity equal to $Z_F$ at the Fermi surface. \\

\section{Discussion}
\label{sec:discussion}

Our findings may apply to the description of multiple semiconductors such as $\alpha-$Sn [\cite{graytinstruct,graytinarpes}], some zinc-blende semiconductors like HgSe and HgTe [\cite{hgtese}], half-Heuslers like YPtBi and LuPtBi [\cite{hfheuslerabinit,hfheuslerabinit2}] and pyrochlores like Pr$_2$Ir$_2$O$_7$ [\cite{pr2ir2o7th,iridatespiniceexp1,iridatespiniceexp2}]. In these materials the average dielectric permittivity is around $\epsilon^{*} \approx 15-20$ [\cite{graytinopt,hgseopt,hgtebosc,hgteeps,hfheuslercalceps,pr2ir2o7}] with an effective mass of $m^*/m_e = 0.2$ for $\alpha-$Sn [\cite{grtinm1,grtinm2}], $0.03-0.07$ for HgSe [\cite{hgsem1,hgsem2}], $0.04$ for HgTe [\cite{hgtebosc}], $0.11-0.23$ for YPtBi and LuPtBi [\cite{yptbim1,yptbim2,luptbimr}], and $6.3$ for Pr$_2$Ir$_2$O$_7$ [\cite{pr2ir2o7,iridatespiniceexp1,iridatespiniceexp2}]. These materials also have similar carrier densities with $n \approx 10^{18}$ cm$^{-3}$ on average [\cite{grtinm2,hgsem1,hgsem2,hgtebosc,yptbim1,yptbim2,luptbimr}] and thus rather similar Wigner-Seitz radii $r_s \approx 0.5-1$. A value as large as $r_s = 10-15$ occurs in the pyrochlore Pr$_2$Ir$_2$O$_7$ due to the large effective mass [\cite{pr2ir2o7}]. This value for Pr$_2$Ir$_2$O$_7$ also corroborates with the zero-temperature residual dielectric permittivity in Ref.~[\cite{pr2ir2o7}] that the authors associate with interband coupling and that we evaluate to $\epsilon_{\rm inter}(\omega = 0, q = 0)/\epsilon^* = 4\alpha r_s/\pi$ (see Appendix~\ref{app:plasma}). This very large value of $r_s$ is not expected for metals where the compressibility is negative beyond $r_s \sim 5$ while it is still positive for Luttinger semimetals (see Fig.~\ref{fig:oneparticle}(c)).

In these materials, we expect a pronounced transmission of light for a frequency $\omega = \omega_T \approx 1.113 E_F$, above the plasma frequency and below the onset of interband transitions (see Fig.~\ref{fig:wprs}(b)). The $1.113$ multiplicative constant changes in the absence of particle-hole symmetry (\emph{i.e.} for $\alpha_1 \neq 0$ in Eq.~\eqref{eq:h0}) and in Appendix~\ref{app:plasma} we illustrate its behaviour, in units of the optical gap. In the case of YPtBi, where ab-initio calculations indicate particle-hole symmetric bands [\cite{brydon}] and with typically $E_F \approx -100$ meV [\cite{yptbim1,yptbim2}], we expect this transmission peak to happen for a light frequency of about $113$ meV. Because this transparency is due to compensating intraband and interband contributions, it can also happen for other band-structures such as Kane semimetals and may be at the origin of the transmission peak observed in Cd$_3$As$_2$ [\cite{cd3as2sp}].

Also, the decrease in the effective mass for increasing values of $r_s$, or decreasing carrier density, illustrated in Fig.~\ref{fig:oneparticle}(b), can be captured through transport measurements. For this purpose, one can use the following fitting expression for the effective mass of the interacting Luttinger model
\begin{align}
    m^*/m \approx 0.861 + 0.126 \exp(-r_s/0.98),
\end{align}
along with the expression $n(r_s) = (3/4\pi) \left(m e^2/(\alpha\epsilon^* r_s)\right)^3$ for carrier density, where $m$ is the bare band mass and $\epsilon^*$ the background dielectric permittivity. We illustrate this in the case of HgSe where the observed carrier density spans over four orders of magnitude, with $n \in [10^{16},10^{19}]$ cm$^{-3}$ [\cite{hgsem2}]. In the low carrier regime we obtain a satisfactory fit to the experimental values of $m^*$ for a bare band mass $m/m_e = 0.0353$ and using $\epsilon^* = 20$ [\cite{hgseopt}]. However, this behaviour does not extend up to higher carrier densities because of the presence of non-parabolic terms in the band structure [\cite{hgsem2}], whose contribution to the effective mass is only negligible at low carrier densities.

Finally, we would like to mention that the experimental observation of the plasmaron mode was reported for other band structure via optical and angle-resolved spectroscopy [\cite{plasmaronbismuth,plasmaronlasing,grapheneplasmaron1,grapheneplasmaron2}]. Its observation for narrow gap semiconductors may be rendered difficult due to the frequent presence of surface states, like in HgTe [\cite{arpeshgte}] and YPtBi [\cite{arpesyptbi}].

\section{Conclusion}
\label{sec:conclusion}

We have derived an analytical expression of the dielectric permittivity of a Luttinger semimetal in the random phase approximation. We have used this expression to compute the single-particle properties of Luttinger's model. The competing intraband and interband transitions are at the origin of a transparency window of the electron gas and a reduced plasma frequency at large $r_s$. The single-particle properties are also affected by interband coupling because of the increased long-range screening, where the bare Coulomb potential is the strongest. The effective mass decreases for all $r_s$ and compressibility shows an instability only for very large values of the Wigner-Seitz radius. This increased stability of the electron gas in Luttinger's model may affect the appearance of correlated phenomena such as Wigner crystallisation and superconductivity [\cite{takada}]. We have also computed the single-particle spectral function which shows plasmaron branches that could be observed in optics and in angle-resolved spectroscopy.

A potential extension of this work is in the evaluation of spin polarizability of Luttinger's semimetals, which expression resembles that of charge polarizability $\Pi$ in Eq.~\eqref{eq:polarizability} up to a form factor. The spin polarizabilities explicitly appear in the Ruderman-Kittel-Kasuya-Yosida (RKKY) interaction between magnetic impurities [\cite{rkkyintro}] and were computed for the surface states of topological insulators [\cite{rkkyti}], quadratic bands with Rashba spin-orbit coupling [\cite{rkky0}], and for Weyl semimetals [\cite{rkky1,rkky2}]. The dependence of the RKKY interaction on the helical structure of the eigenstates may help further characterize Luttinger semimetals through their magnetic properties. Moreover, in the case of Pr$_2$Ir$_2$O$_7$, this interaction may be responsible for the magnetic coupling between the Pr local moments [\cite{iridatespiniceth}].

%In this work, the effective potential between electrons only includes the screened potential. While this might be valid for a test charge, the electron gas should also include exchange and correlation interactions described by vertex and electron-hole ladder corrections. This can be related to the spin polarizabilities $\Pi_{ij}$ [\cite{spinpol}] that lead to additional spin modes and that are responsible for magnetic coupling in the RKKY interaction [\onlinecite{rkky1,rkky2}]. 
~~\\
\begin{acknowledgments}
We would like to thank \'Eric Dupuis for fruitful discussions.  
ST and WWK were funded by a Discovery Grant from NSERC, a Canada Research Chair, and a ``\'Etablissement de nouveaux chercheurs et de nouvelles chercheuses universitaires'' grant from FRQNT. This research was enabled in part by support provided by Calcul Qu\'ebec (www.calculquebec.ca) and Compute Canada (www.computecanada.ca).
\end{acknowledgments} 

\bibliographystyle{apsrev4-1}
\bibliography{bibliography}

\newpage

\appendix
\begin{widetext}
\section{Charge polarizability}
\label{app:pintrainter}

We can decompose the expression of the polarizability in two parts
\begin{align}
    \Pi(i\Omega,{\bf q}) &= \sum_{\sigma \sigma^{\prime} \bf p} \frac{f_{D}(\varepsilon_{\sigma}({\bf p})) - f_{D}(\varepsilon_{\sigma^{\prime}}({\bf p}+{\bf q}))}{i\Omega +\varepsilon_{\sigma}({\bf p}) - \varepsilon_{\sigma^{\prime}}({\bf p}+{\bf q})} {\rm Tr}\left[\hat{P}_{\sigma}({\bf p}) \hat{P}_{\sigma^{\prime}}({\bf p} +{\bf q})\right] \\
    &= \sum_{\sigma \sigma^{\prime} \bf p} \frac{f_{D}(\varepsilon_{\sigma}({\bf p}))}{i\Omega +\varepsilon_{\sigma}({\bf p}) - \varepsilon_{\sigma^{\prime}}({\bf p}+{\bf q})} {\rm Tr}\left[\hat{P}_{\sigma}({\bf p}) \hat{P}_{\sigma^{\prime}}({\bf p} +{\bf q})\right] + i\Omega \leftrightarrow -i\Omega
\end{align}
where the trace of the product of projectors on each subband $\sigma = \pm$ is
\begin{align}
    {\rm Tr}\left[\hat{P}_{\sigma}({\bf p}) \hat{P}_{\sigma^{\prime}}({\bf p} +{\bf q})\right] = \frac12\left\{ 2 + \sigma\sigma^{\prime} \left[ 3 \cos^2(\theta_{{\bf p}, {\bf p} + {\bf q}}) - 1\right] \right\}. 
\end{align}
Then we have, writing the polarizability in units of the density of states at the Fermi energy $N_0 = 1/(4\pi^2)$, the wavevectors in units of $k_F$ and the frequencies in units of $E_F$, 
\begin{align}
    \Pi(i\Omega,{\bf q}) &= \frac{N_0}{2} \sum_{\sigma \sigma^{\prime}} \int dp~ p^2 d\theta_{{\bf p},{\bf p}+{\bf q}} \sin(\theta_{{\bf p},{\bf p}+{\bf q}}) \frac{f_{D}(\varepsilon_{\sigma}({\bf p}))}{i\Omega +\varepsilon_{\sigma}({\bf p}) - \varepsilon_{\sigma^{\prime}}({\bf p}+{\bf q})} \left( 2 + \sigma\sigma^{\prime}\left[3\cos^2(\theta_{{\bf p},{\bf p}+{\bf q}}) - 1 \right] \right) + i\Omega \leftrightarrow -i\Omega\\
    &= \frac{N_0}{2} \sum_{\sigma \sigma^{\prime}} \int dp~ p^2 du \frac{f_{D}(\varepsilon_{\sigma}({p}))}{i\Omega  + (\sigma - \sigma^{\prime}) p^2 - \sigma^{\prime}(q^2 + 2 pq u)} \left( \delta_{\sigma\sigma^{\prime}} 4 + \sigma\sigma^{\prime}\frac{3(u^2-1)q^2}{q^2 + p^2 + 2 p q u} \right) + i\Omega \leftrightarrow -i\Omega\\
    &= \Pi^{\rm intra}(i\Omega,{\bf q}) + \Pi^{\rm inter}(i\Omega,{\bf q})
\end{align}
where we introduce the intraband and interband contributions to the polarizability
\begin{align}
    \Pi^{\rm intra}(i\Omega,{\bf q}) &= \frac{N_0}{2} \int_{1}^{\infty} dp~ p^2~ \int_{-1}^{1} du~ \frac{1}{i\Omega + q^2 + 2 pq u} \left( 4 + \frac{3(u^2-1)q^2}{q^2 + p^2 + 2 p q u} \right) + i\Omega \leftrightarrow -i\Omega,\\
    \Pi^{\rm inter}(i\Omega,{\bf q}) &= - \frac{N_0}{2} \int_{1}^{\infty} dp~ p^2~ \int_{-1}^{1} du~ \frac{1}{i\Omega  - 2 p^2 - q^2 - 2 pq u} \frac{3(u^2-1)q^2}{q^2 + p^2 + 2 p q u} + i\Omega \leftrightarrow -i\Omega.
\end{align}

\subsection{Intraband contribution}
Since the Fermi energy is in the bottom band, with spectrum $\epsilon_{-}(k) = -k^2/2m$, we have integrated the polarizability for $p \in [1,\infty[$. Up to an overall negative sign, one can perform the integration for $p \in [0,1]$ since the intraband contribution vanishes for a completely filled band (\emph{i.e.} when $p \in [0,\infty[$). We decompose the intraband polarizability in a component that corresponds to a single quadratic band, described by the Lindhard polarizability, and a correction due to the eigenspinor overlap
\begin{align}
    \Pi^{\rm intra}(i\Omega,{\bf q}) &= 2 \Pi_{\rm Lindhard}(i\Omega,{\bf q}) + \Pi_{\rm overlap}(i\Omega,{\bf q})
\end{align}
where
\begin{align}
    \Pi_{\rm Lindhard}(i\Omega,{\bf q}) &= - N_0\int_{0}^{1} dp~ p^2~ \int_{-1}^{1} du~ \frac{1}{i\Omega + q^2 + 2 pq u}  + i\Omega \leftrightarrow -i\Omega\\
    \Pi_{\rm overlap}(i\Omega,{\bf q}) &= - \frac{N_0}{2} \int_{0}^{1} dp~ p^2~ \int_{-1}^{1} du~ \frac{1}{i\Omega + q^2 + 2 pq u} \frac{3(u^2-1)q^2}{q^2 + p^2 + 2 p q u} + i\Omega \leftrightarrow -i\Omega
\end{align}
The Lindhard polarizability can be found in numerous works [\cite{giuliani}] :
\begin{align}
    \Pi_{\rm Lindhard}(i\Omega,{\bf q}) &= -\frac{N_0}{q}\left[ \Psi_{\rm Lindhard}\left(\frac{i\Omega}{2q} - \frac{q}{2}\right) - \Psi_{\rm Lindhard}\left(\frac{i\Omega}{2q} + \frac{q}{2}\right) \right]
\end{align}
with the Lindhard function
\begin{align}
    \Psi_{\rm Lindhard}(z) = \frac12\left[ z + \frac{1-z^2}{2}\log\left( \frac{z+1}{z-1} \right) \right].
\end{align}
We compute the overlap contribution in two steps. First, we perform the integration over angles after a partial fraction decomposition 
\begin{align}
    &\int_{-1}^{1} du~ \frac{p^2}{i\Omega + q^2 + 2 pq u} \frac{3(u^2-1)q^2}{q^2 + p^2 + 2 p q u} \\
    &= \frac{3}{2q} \int_{-1}^{1} du~ \left(\frac{q}{2} + \frac{1}{ 2q } \frac{(p^2-q^2)^2}{i \Omega - p^2}\frac{1}{ 1 + (p/q)^2 + 2pu/q } -q^2 \frac{p^2-(i\Omega/2q+q/2)^2}{p^2-i\Omega}\frac{1}{(i\Omega/2q+q/2) + p u}\right)\\
    &= \frac{3}{2q} \left(q + \frac{1}{ 2q } \frac{(p^2-q^2)^2}{i \Omega - p^2} \frac{q}{p}\log\left( \frac{ 1 + p/q }{|1 - p/q|} \right) -q^2 \frac{p^2-(i\Omega/2q+q/2)^2}{p^2-i\Omega}\frac1p \left[ \log\left( i\Omega/2q+q/2 + p \right) - \log\left( i\Omega/2q+q/2 - p \right)\right]\right)
    \label{eq:apppolaintra}
\end{align}
The integration over $p$ can then be expanded as a sum of integrals of the form
\begin{align}\label{eq:defIpm}
    I_{+}(a,b ; \alpha,\beta) &= \int_{a}^{b}dx~ \frac{\log(x+\alpha)}{x+\beta},\\
    I_{-}(a,b ; \alpha,\beta) &= \int_{a}^{b}dx~ \frac{\log(|x-\alpha|)}{x+\beta},
\end{align}
which were discussed in the context of hole screening in zinc-blende semiconductors [\cite{zcbldhole}] and loop integrals in quark physics [\cite{loopintegral}]. These expressions resemble the Spence function but where one has to account for the singularity at $x = -\beta$ and the discontinuity of the logarithm on the negative real axis. We use Ref.~[\cite{loopintegral}] and find
\begin{align}
    \label{eq:Ipmexp}
    I_{+}(a,b ; \alpha,\beta) &= \left[ {\rm Li}_2\left(1 - \frac{\alpha + a}{\alpha - \beta}\right) + \eta\left(\alpha + a, 1/(\alpha - \beta)\right)\log(1 - \frac{\alpha + a}{\alpha - \beta}) - \{a \rightarrow b\} \right] + \log(\alpha - \beta)\log\left[ \frac{b + \beta}{a + \beta} \right]\\
    I_{-}(a,b ; \alpha,\beta) &= \left\{
    \begin{array}{lr}
        \left[ {\rm Li}_2\left(1 - \frac{\alpha - a}{\alpha + \beta}\right) + \eta\left(\alpha - a, 1/(\alpha + \beta)\right)\log(1 - \frac{\alpha - a}{\alpha + \beta}) - \{a \rightarrow b\} \right] + \log(\alpha + \beta)\log\left[ \frac{b + \beta}{a + \beta} \right] & \textrm{if $\alpha > b$}\\
        {\rm Li}_2\left(1 - \frac{\alpha - a}{\alpha + \beta}\right) - \log(\alpha + \beta)\log\left[ \frac{a + \beta}{\alpha + \beta} \right] - \{a,\alpha,\beta \rightarrow -b,-\alpha,-\beta\}  & \textrm{if $\alpha \leq b$}
    \end{array}
    \right.
\end{align}
where for simplicity we write $I_{-}$ in the case where $a$, $\alpha$ and $b$ are all real, because this is the situation at hand. We have also introduced the dilogarithm ${\rm Li}_2(z) = - \int_{0}^{z} du~ \ln(1-u)/u$ with $z \in \mathbb{C}$ and the function $\eta(a,b)$ that keeps track of the discontinuity in the logarithm, such that $\log(ab) = \log(a) + \log(b) + \eta(a,b)$,
\begin{align}
    \eta(a,b) = \left\{
        \begin{array}{ll}
            2i\pi & \textrm{if ${\rm Im}(a) < 0$, ${\rm Im}(b) < 0$ and ${\rm Im}(ab) > 0$,}\\
            -2i\pi & \textrm{if ${\rm Im}(a) > 0$, ${\rm Im}(b) > 0$ and ${\rm Im}(ab) < 0$ or if ${\rm Im}(a) = {\rm Im}(b) = 0$, ${\rm Re}(a) < 0$ and ${\rm Re}(b) < 0$,}\\
            0 & \textrm{else.}
        \end{array}
    \right.
\end{align}
We thus find the following expression for the overlap contribution to intraband polarizability
\begin{align}
    \Pi_{\rm overlap}(i\Omega,{\bf q}) &= \Psi_2\left(i\Omega,{\bf q}\right) + \{i\Omega \rightarrow - i \Omega \},
\end{align}
with $\Psi_2(i\Omega,{\bf q})$ defined in Eq.~\eqref{eq:psi2} of the main text.

\subsection{Interband contribution}

The interband contribution to the polarizability is treated in a similar fashion. We first perform the integration over angles after a partial fraction decomposition 
\begin{align}
    &-\int_{-1}^{1} du~ \frac{p^2}{i\Omega  - 2 p^2 - q^2 - 2 pq u} \frac{3(u^2-1)q^2}{q^2 + p^2 + 2 p q u}\\
    &= 3\int_{-1}^{1} du~ \left\{ \left[ \frac{(p^2-q^2)^2}{4q^2(p^2 - i\Omega)}\frac{1}{1+(p/q)^2-2(p/q)u} - \frac14 \right] + \left[ \frac{4p^2(p^2 - i\Omega) + (q^2 - i\Omega)^2}{4(p^2-i\Omega)} \frac{1}{i\Omega - q^2 - 2p^2 + 2p q u} + \frac{1}{2} \right] \right\}\\
    &= 3\left\{ \frac{1}{4q} \left[ \left( p - \frac{q^4}{i\Omega p} + \frac{(q^2 - i\Omega)^2}{i\Omega}\frac{p}{p^2 - i\Omega} \right)\log\left( \frac{1+p/q}{|1-p/q|}\right) - 2q \right] \right.\\
    &\left.+ \left[ 1 + \frac{1}{8q} \left(4 p + \frac{(q^2 - i\Omega)^2}{2i\Omega} \frac{2 p }{p^2 - i\Omega} - \frac{(q^2-i\Omega)^2}{i\Omega p} \right) \left( \log\left( i\Omega - q^2 - 2p^2 + 2 p q \right) - \log\left( i\Omega - q^2 - 2p^2 - 2 p q \right) \right)\right] \right\}\nonumber
\end{align}
In the last line, each square brackets is convergent when integrated over $p$. This integral is similar to that for the intraband polarizability in Eq.~\eqref{eq:apppolaintra} with two differences : ($i$) one can simplify the integration domain with the change of variable $p \rightarrow 1/p$ and, ($ii$) some of the logarithms contain a second order polynomial that has to be decomposed into a product of monomials to recover the integrals in Eq.~\eqref{eq:defIpm}. We get :
\begin{align}
    \Pi^{\rm inter}(i\Omega,{\bf q}) = \Psi_3(i\Omega,{\bf q}) + \Psi_3(-i\Omega,{\bf q})
\end{align}
with $\Psi_3(i\Omega,{\bf q})$ defined in Eq.~\eqref{eq:psi3} of the main text.

\section{Plasma frequency}
\label{app:plasma}

The small $q$ expansion of the dielectric permittivity writes 
\begin{align}
    \epsilon(\omega)  = \lim_{q\rightarrow 0} \epsilon(\omega,{\bf q}) = 1 - \lim_{q\rightarrow 0} V({\bf q}) \Pi(\omega,{\bf q}),
\end{align}
with the following intraband and interband contributions
\begin{align}
    \lim_{q\rightarrow 0} V(q) \Pi^{\rm intra}(\omega,{\bf q}) &= \frac{16\alpha r_s}{3\pi \omega^2},\\
    \lim_{q\rightarrow 0} V(q) \Pi^{\rm inter}(\omega,{\bf q}) &= \frac{\alpha r_s}{\pi \sqrt{\omega/2}} \left( \log\left( \frac{1-\sqrt{(\omega+i0^{+})/2}}{1+\sqrt{(\omega+i0^{+})/2}} \right) - 2 \arctan(\sqrt{\omega/2}) \right).
\end{align}
Note that in the expression of the interband contribution, we explicitly keep $\omega + i0^+$ due to the analytic continuation from complex frequencies. The same expressions for the optical response of Luttinger semimetals were obtained in Refs.~[\cite{q01,w05}]. We have numerically determined that these two expressions cancel each other for $\omega_T/E_F \approx 1.113$, which we call the transparency frequency in the main text.

The equation for the plasma frequency is $\epsilon(\omega = \omega_p) = 0$ which is transcendental in presence of the interband contribution. We have solved it numerically in Fig.~\ref{fig:wprs} and one can find an approximate expression since $\lim_{q\rightarrow 0} V(q) \Pi^{\rm inter}(\omega,{\bf q}) \approx -4\alpha r_s/\pi$ for small frequencies. This term gives a large contribution to the imaginary part of the optical conductivity at low frequency and low temperatures, as observed in Pr$_2$Ir$_2$O$_7$ [\cite{pr2ir2o7}]. With this small energy expansion, we find the approximate expression for the plasma frequency
\begin{align}
    \omega_p \approx \sqrt{\frac{16 \alpha r_s}{3\pi(1+4\alpha r_s/\pi)}}.
\end{align}

We have also derived these expressions in the case of non particle-hole symmetric bands, for $\alpha_1 \neq 0$ in Eq.~\eqref{eq:h0}. The interband polarizability $\Pi^{\rm inter}$ only depends on the difference in energy between the two bands and thus only on $\alpha_2$ while the intraband contribution depends on the mass of the filled band, $m_{\pm} = m/(\alpha_2 \pm \alpha_1) > 0$ (\emph{i.e.} $|\alpha_1| < |\alpha_2|$). Then the two expressions are the same if we change the energy scale from $E_F = \hbar^2 k_F^2/2m$ to that of the optical gap $2E_0 = \hbar^2 k_F^2(1/2m_+ + 1/2m_-) =  \alpha_2 \hbar^2 k_F^2/m$, which affects our definition of $r_s$ to $m e^2/(\alpha \alpha_2\epsilon^* k_F)$, and if one multiplies $\lim_{q\rightarrow 0} V(q) \Pi^{\rm intra}(\omega,{\bf q})$ by $\beta = 1 \pm \alpha_1/\alpha_2 \in [0 ; 2]$ with $\pm$ for the Fermi energy in the upper or lower band. If $\beta = 0$ then the band at the Fermi surface is flat and, on the contrary, if $\beta = 2$ then the other band is flat. The transparency frequency in units of $E_0$ then behaves as in Fig.~\ref{fig:wtr} as a function of $\beta$. We recover $\omega_{T}(\beta = 1)/E_0 = 1.113$ as discussed in the main text and also that $\omega_{T}(\beta = 0) = 0$ for a Fermi surface at a flat band and $\omega_{T}(\beta = 2)/E_0 = 1.506$ for a Fermi surface at the light band touching a flat band. We can also derive the following approximate expression of the plasma frequency in units of $E_0$
\begin{align}
    \omega_p(\beta) \approx \sqrt{\frac{16 \alpha r_s \beta}{3\pi(1+4\alpha r_s/\pi)}}.
\end{align}

\begin{figure}[t!]
    \centering
    \includegraphics[width = 0.45\textwidth]{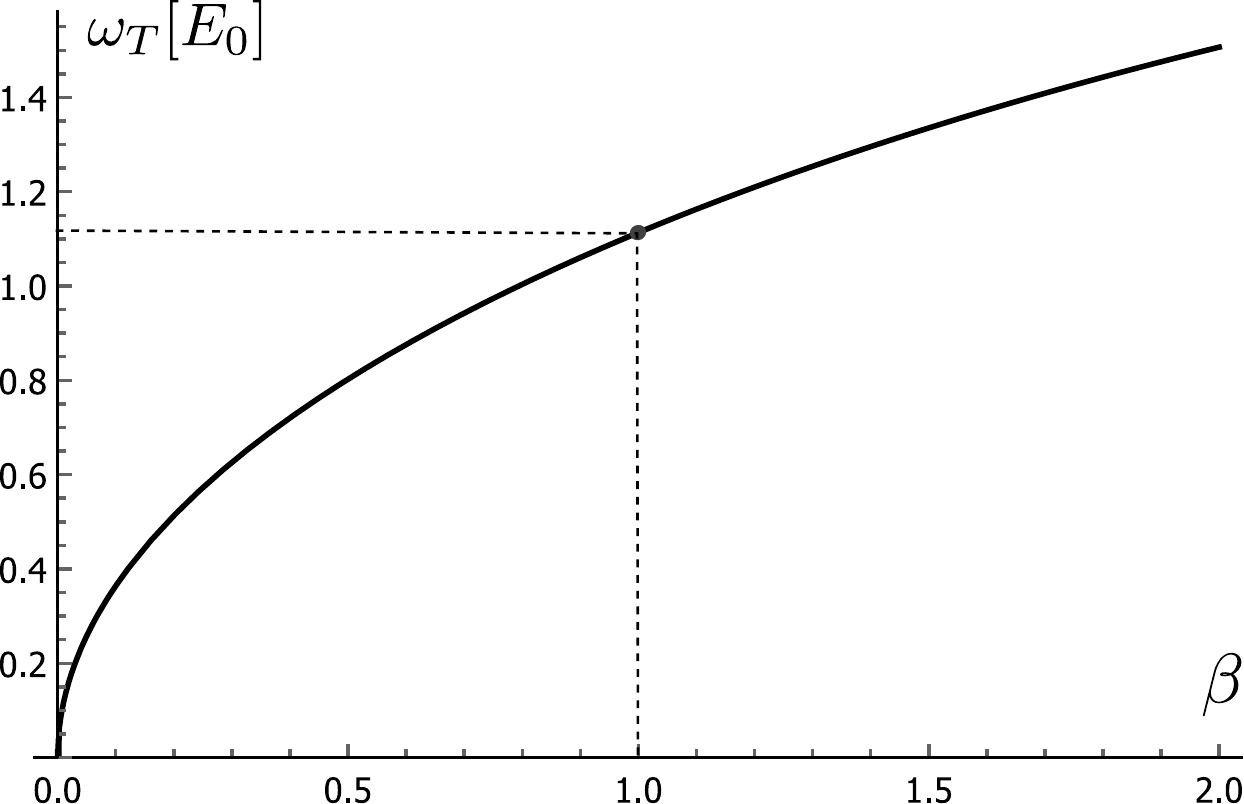}
    \caption{Behaviour of the transparency frequency $\omega_T$, such that the electronic polarizability vanishes $\lim_{q\rightarrow 0}\Pi(q,\omega_T) = 0$, for non particle-hole symmetric bands. The band at the Fermi surface is flat for $\beta = 1 - \alpha_1/\alpha_2 = 0$ and the other one for $\beta = 2$. The dashed line indicates the situation of particle-hole symmetry, where $\beta = 1$.}
    \label{fig:wtr}
\end{figure}

\section{Charge-spin response}
\label{app:wpnospin}

As discussed in the main text, the eigenstates of Luttinger's model Eq.~\eqref{eq:h0} can be associated to eigenstates of the helicity operator $\hat{\lambda} = {\bf k}\cdot \hat{{\bf J}}/k$ [\cite{schliemann,savary}]. A similar helical structure is associated to a charge-spin response in Weyl semimetals because of spin-momentum locking and is responsible for spin polarized plasmon excitations [\cite{chargespin}]. Here, in the case of a quadratic band touching, the charge-spin response identically vanishes due to the compensating helicities of each band.

We can explicitly compute the charge-spin response for the quadratic band touching, defined by
\begin{align}
    \Pi_{\rho \hat{J}_0}(i\Omega,{\bf q}) &= \sum_{\sigma \sigma^{\prime} \bf p} \frac{f_{D}(\varepsilon_{\sigma}({\bf p})) - f_{D}(\varepsilon_{\sigma^{\prime}}({\bf p}+{\bf q}))}{i\Omega +\varepsilon_{\sigma}({\bf p}) - \varepsilon_{\sigma^{\prime}}({\bf p}+{\bf q})} {\rm Tr}\left[\hat{P}_{\sigma}({\bf p}) \hat{\bf J}\cdot {\bf e}_0 \hat{P}_{\sigma^{\prime}}({\bf p} +{\bf q})\right],
\end{align}
where $\hat{J}_{0} = \hat{\bf J}\cdot {\bf e}_0$ is a component of the spin-$3/2$ operator in the direction ${\bf e}_0$ and
\begin{align}
    {\rm Tr}\left[\hat{P}_{\sigma_2}({\bf k}_2) \hat{\bf J}\cdot {\bf e}_0 \hat{P}_{\sigma_1}({\bf k}_1)\right] = \frac{3i\sigma_1\sigma_2}{2k_1^2 k_2^2} {\bf k}_1\cdot {\bf k}_2 \left( {\bf k}_1\times {\bf k}_2 \right)\cdot{\bf e}_0.
\end{align}
In as similar manner to Weyl semimetals [\cite{chargespin}], the expression $\Pi_{\rho \hat{J}_0}({\bf q},i\Omega)$ vanishes for ${\bf q} \perp {\bf e}_0$ because the spectrum is even when changing the sign of any component of ${\bf k}$ while the form factor is odd for the component transverse to both ${\bf q}$ and ${\bf e}_0$. And, unlike for Weyl semimetals, in the situation ${\bf q} \parallel {\bf e}_0$ the form factor also vanishes because of the cross product. Thus, for any direction ${\bf q}$ one finds 
\begin{align}
    \Pi_{\rho \hat{J}_0}(i\Omega,{\bf q}) = 0.
\end{align}
The charge and the spin excitations are thus decoupled and plasmons do not carry spin but only charge. In the more general Luttinger model [\cite{luttinger}], lifting the degeneracy on each subband may allow for a non-zero charge-spin coupling. This was for example discussed for Weyl nodes with opposite helicites but with different band dispersion [\cite{chargespin}] and for 2D electron gases in presence of Rashba spin-orbit coupling [\cite{spinpol}].

\section{Self-energy}

The calculation of the self-energy follows the $GW$ approximation
\begin{align}
    \Sigma_{\sigma}(i\nu,{\bf q}) &= -\frac{1}{2\mathcal{V}}\beta^{-1}\sum_{\Omega \sigma^{\prime} {\bf k} } V({\bf k})\left( 1 + \frac{1}{\epsilon(i\Omega,{\bf k})} - 1 \right) {\rm Tr}\left[\hat{P}_{\sigma}({\bf q}) \hat{P}_{\sigma^{\prime}}({\bf q -k})\right] G_{\sigma^{\prime}}(i(\nu - \Omega),{\bf q}-{\bf k})\\
    &= \Sigma_{\sigma}^{\rm (ex)}({\bf q}) + \Sigma_{\sigma}^{\rm (c)}(i\nu,{\bf q})
\end{align}
where we introduce the exchange and correlation self-energies
\begin{align}
    \Sigma_{\sigma}^{\rm (ex)}({\bf q}) &= -\frac{1}{2\mathcal{V}}\beta^{-1}\sum_{\Omega \sigma^{\prime} {\bf k} } V({\bf k}) {\rm Tr}\left[\hat{P}_{\sigma}({\bf q}) \hat{P}_{\sigma^{\prime}}({\bf q - k})\right] G_{\sigma^{\prime}}(i(\nu - \Omega),{\bf q}-{\bf k})\\
    &= - \frac{1}{2\mathcal{V}}\sum_{\sigma^{\prime} {\bf k} } f_{D}(\xi_{\sigma^{\prime}{\bf q}-{\bf k}}) V({\bf k}) {\rm Tr}\left[\hat{P}_{\sigma}({\bf q}) \hat{P}_{\sigma^{\prime}}({\bf q - k})\right]\\
    &= - \frac{1}{2\mathcal{V}}\sum_{\sigma^{\prime} {\bf k} } (f_{D}(\xi_{\sigma^{\prime}{\bf k}}) - \tilde{f}_{\sigma^{\prime},k} + \tilde{f}_{\sigma^{\prime},k})  V({\bf q}-{\bf k}) {\rm Tr}\left[\hat{P}_{\sigma}({\bf q}) \hat{P}_{\sigma^{\prime}}({\bf k})\right] = \Sigma_{\sigma}^{int}({\bf q}) + \Sigma_{\sigma}^{ext}({\bf q}),\\
    \Sigma_{\sigma}^{\rm (c)}(i\nu,{\bf q}) &= -\frac{1}{2\mathcal{V}} \beta^{-1}\sum_{\Omega \sigma^{\prime} {\bf k} } V({\bf k})\left( \frac{1}{\epsilon(i\Omega,{\bf k})} - 1 \right) {\rm Tr}\left[\hat{P}_{\sigma}({\bf q}) \hat{P}_{\sigma^{\prime}}({\bf q -k})\right] G_{\sigma^{\prime}}(i(\nu - \Omega),{\bf q}-{\bf k}).
\end{align}

\subsection{Exchange self-energy}
\label{sec:exse}

We decompose the exchange self-energy in a intrisinc and an extrinsic part, $\Sigma^{\rm (ex)} = \Sigma_{\sigma}^{\rm int} + \Sigma_{\sigma}^{\rm ext}$. In the intrinsic contribution the situation is as if the chemical potential is at the quadratic band touching
\begin{align}
    \Sigma_{\sigma}^{\rm int}({\bf k}) &= - \frac{\alpha r_s}{\pi} \int_{0}^{k_c} dq~ q^2 \int_{-1}^{1} du~ \frac{1}{k^2 + q^2 - 2 k q u} {\rm Tr}\left[\hat{P}_{\sigma}({\bf k}) \hat{P}_{-}({\bf q})\right].
\end{align}
The integral does not converge for large wavevectors and we have introduced a momentum cutoff $\Lambda \equiv k_c k_F$ (e.g. due to the lattice constant), with $k_c$ the corresponding dimensionless quantity, on which the final expression depends 
\begin{align}
    \Sigma_{\sigma}^{\rm int}({\bf k}) &= - \frac{\alpha r_s k}{\pi} \int_{0}^{k_c/k} dx~ \left( \frac{3 \sigma}{4} \left( 1 + x^2 \right) + \left( x - \sigma \frac{3+2x^2+ 3x^4}{8x} \right)\log\left(\frac{1+x}{|1-x|}\right) \right)\\
    &= - \frac{\alpha r_s k}{\pi} \left( \left[ 1 + \frac{\sigma}{16}\left( 5 + 3 \frac{k_c^2}{k^2} \right) \right]\frac{k_c}{k} + \frac{(k_c^2-k^2)(16 k^2 - \sigma(7k^2 + 3k_c^2) )}{32k^4} \log\left( \frac{1+k/k_c}{|1-k/k_c|} \right) \right.\\
    &\left. - \frac{\sigma}{16}\left[ \pi^2 - 6 \log\left(\left|1-\frac{k_c}{k}\right|\right)\log\left(\frac{k_c}{k}\right) - 6 \left( {\rm Li}_2\left(1-\frac{k_c}{k}\right) + {\rm Li}_2\left(-\frac{k_c}{k}\right)\right) \right]  \right)\\
    &\approx_{k_c \gg 1} - \frac{\alpha r_s k}{\pi}  \left( 2 \frac{k_c}{k} - \frac{3\pi^2}{16} \sigma \right).
\end{align}
On the other hand, the extrinsic contribution to the exchange self-energy does not need cutoff and has a similar angular integration than the intrinsic part. Its expression is
\begin{align}
    \Sigma_{\sigma}^{\rm ext}({\bf k}) &= \frac{\alpha r_s}{\pi} \int_{0}^{1} dq~ q^2 \int_{-1}^{1} du~ \frac{1}{k^2 + q^2 - 2 k q u} {\rm Tr}\left[\hat{P}_{\sigma}({\bf k}) \hat{P}_{-}({\bf q})\right]\\
    &= \frac{\alpha r_s}{\pi} \left( \frac12(2+\sigma) + \frac{3}{16}\sigma(1/k^2 -1) + \frac{(1-k^2)(16 k^2 - \sigma(7k^2 + 3) )}{32k^3} \log\left( \frac{1+k}{|1-k|} \right) \right.\\
    &\left. - \frac{\sigma k}{16}\left[ \pi^2 + 6 \log\left(\left|1-k\right|\right)\log\left({k}\right) - 6 \left( \log^2(k) + {\rm Li}_2\left(1-\frac{1}{k}\right) + {\rm Li}_2\left(-\frac{1}{k}\right)\right) \right]  \right).
\end{align}

We then obtain, in the limit $k_c = \Lambda/k_F \gg 1$, the following expression for the exchange self-energy
\begin{align}
     \Sigma_{\sigma}^{\rm (ex)}({\bf q}) \approx \frac{\alpha r_s}{\pi} &\left\{ -2 \Lambda/k_F + \frac{3\sigma\pi^2 q}{16} + \frac{1}{32} \left[ 16(2+\sigma) + 6\sigma\left( 1/q^2 - 1 \right) + \frac{(q^2-1)(3\sigma + q^2(7\sigma - 16))}{q^3} \log\left( \frac{1+q}{|1-q|} \right)\right.\right.\\
     &\left.\left. - 2\sigma q\left( \pi^2  + 6\left\{ \log\left( |1-q| \right)\log( q ) - \log^2(q) - {\rm Li}_{2}(-1/q) - {\rm Li}_{2}(1-1/q) \right\} \right) \right]\right\}.
\end{align}
The self-energy explicitly depends on the cutoff wavevector, $\Lambda$, but only like a constant shift that one can absorb in the definition of the chemical potential. 

\subsection{Correlation self-energy}
We perform the analytical continution of the Matsubara frequency to the real frequency in the correlation self-energy. This leads to the following expression for the correlation energy in real frequencies
\begin{align}
    \Sigma_{\sigma}^{\rm (c)}(\omega,{\bf q}) &= \Sigma_{{\rm line},\sigma}(\omega, k) + \Sigma_{{\rm res},\sigma}(\omega, k),
\end{align}
with
\begin{align}
    \Sigma_{{\rm line},\sigma}(\omega, k) &= - \frac{1}{2\mathcal{V}}\sum_{\sigma^{\prime} {\bf q}} \int\frac{d\Omega}{2\pi}~ G_{\sigma^{\prime}}(\omega - i\Omega, {\bf k} - {\bf q})V({\bf q})\left[ \frac{1}{\epsilon(i\Omega,q)} - 1 \right]{\rm Tr}\left[\hat{P}_{\sigma}({\bf k}) \hat{P}_{\sigma^{\prime}}({\bf k -q})\right], \\
    \Sigma_{{\rm res},\sigma}(\omega, k) &= + \frac{1}{2\mathcal{V}}\sum_{\sigma^{\prime} {\bf q}} \left[ \Theta(\omega - \xi_{\sigma^{\prime}}({\bf k} - {\bf q})) - \Theta( - \xi_{\sigma^{\prime}}({\bf k} - {\bf q})) \right]V({\bf q}) \left[ \frac{1}{\epsilon\left[\omega - \xi_{\sigma^{\prime}}({\bf k} - {\bf q}) - i\eta {\rm sgn}(\omega-\xi_{\sigma^{\prime}}({\bf k-q})),{\bf q}\right]} - 1 \right]\nonumber\\
    &~~~~~~~~~~ \times{\rm Tr}\left[\hat{P}_{\sigma}({\bf k}) \hat{P}_{\sigma^{\prime}}({\bf k -q})\right].
\end{align}

\end{widetext}

\end{document}